\def\rfr#1{eq. (\ref{#1})}

\def\nk{n_{\rm b}}

\def\dert#1#2{\frac{{{d}}{#1}}{{{d}}{#2}}}              
\def\kxy{\mathcal{K}_{xy}}
\def\kxx{\mathcal{K}_{xx}}
\def\kyy{\mathcal{K}_{yy}}
\def\kxz{\mathcal{K}_{xz}}
\def\kyz{\mathcal{K}_{yz}}
\def\Om{\mathit{\Omega}}
\def\Ps{\mathit{\Psi}}
\def\virg#1{``#1''}

\def\eqi{\begin{equation}}
\def\eqf{\end{equation}}
\def\eqia{\begin{eqnarray}}
\def\eqfa{\end{eqnarray}}
\def\rp#1#2{{#1\over#2}} \def\lb#1{\label{#1}}
\def\bb#1#2#3{\bibitem[\protect\citeauthoryear{#1}{#2}]{#3}}

\def\bds#1{\boldsymbol{#1}}

\documentclass{aastex}
\usepackage{url}
\usepackage{amsmath,amsthm,amscd,amssymb}
\usepackage{latexsym}
\usepackage{graphicx,epsfig}

\RequirePackage{color}

\newcommand{\emaila}{lorenzo.iorio@libero.it}

\begin{document}

\title{Orbital effects of a  monochromatic plane gravitational wave with ultra-low frequency incident on a gravitationally bound two-body system}
\shortauthors{L. Iorio}

\author{Lorenzo Iorio\altaffilmark{1} }
\affil{Ministero dell'Istruzione, dell'Universit\`{a} e della Ricerca (M.I.U.R.). Permanent address for correspondence: Viale Unit\`{a} di Italia 68, 70125, Bari (BA), Italy. Tel. +39 329 23 99 167}

\email{\emaila}

\begin{abstract}
We analytically compute the long-term orbital variations of a test particle orbiting a central body acted upon by an incident monochromatic plane gravitational wave. We assume that the characteristic size of the perturbed two-body system is much smaller than the wavelength of the wave. Moreover, we also suppose that the wave's frequency $\nu_g$ is  much smaller than the particle's orbital one $\nk$. We make neither a priori assumptions about the direction of the wavevector {$\bds{\hat{k}}$} nor on the orbital {configuration} of the {particle}.
While the semi-major axis $a$ is left unaffected, the eccentricity $e$, the inclination $I$, the longitude of the ascending node $\Om$, the longitude of pericenter $\varpi$ and the mean anomaly $\mathcal{M}$ undergo non-vanishing long-term changes of the form $d\Ps/dt=\textcolor{black}{F}(\mathcal{K}_{ij};e,I,\Om,\omega),\Ps=e,I,\Om,\varpi,\mathcal{M}$ where $\mathcal{K}_{ij},\ i,j=1,2,3$ are the coefficients of the tidal matrix \textbf{\textsf{K}}. Thus, in addition to the  variations of its orientation in space, the shape of the orbit would be {altered} as well. Strictly speaking, such effects  are not secular trends because of the slow modulation introduced by
\textbf{\textsf{K}} and by the orbital elements themselves: they exhibit peculiar long-term temporal patterns which would be potentially of help for their detection in multidecadal analyses of extended data records of planetary observations of various kinds. In particular, they could be useful in performing independent tests of the inflation-driven ultra-low gravitational waves whose imprint may have been indirectly detected in the Cosmic Microwave Background by the Earth-based experiment BICEP2. Our calculation holds, in general, for any gravitationally bound two-body system whose orbital frequency $\nk$ is much larger than the frequency $\nu_g$ of the external wave, {like, e.g., extrasolar planets and the stars orbiting the Galactic black hole}. It is also valid for a generic perturbation of tidal type with constant coefficients over timescales of the order of the orbital period of the perturbed particle.
 \end{abstract}


\section{Introduction}
Gravitational waves \textcolor{black}{\citep{Flan05,Poi07}} are a key theoretical {prediction}
of the {General Theory of Relativity (GTR)}. Indeed, since {GTR} relies upon the Lorentz invariance,
which carries with it the concept of a limiting speed for {physical} interactions, the existence of gravitational waves is a natural consequence of it.
A direct measurement of them is still lacking; see, e.g., \citet{Cer,Giaz,Fair}  for recent reviews of the status of gravitational wave detection.
To date, only indirect evidences of their existence have been inferred from the orbital decay rate of the binary pulsar PSR B1913+16 \citep{Hul} and, more recently, from the detection \citep{BICEP1} of $B-$mode polarization at degree angular scales in the Cosmic Microwave Background (CMB) by the ground-based experiment BICEP2 \citep{BICEP2} at the South Pole.
The consequences of detecting gravitational waves  for physics, astrophysics, and cosmology would be certainly remarkable; see, e.g., \citet{Satya}.
The role of a direct detection of the gravitational waves for {GTR} and extended theories of gravity was pointed out by \citet{Corda1,Corda2}.

The gravitational wave spectrum  {covers an interval of about}  18 orders of
magnitude in {wavelengths}, encompassing a very broad range of physics and astrophysics \citep{Prince}. The frequencies in the range $10^1-10^4$ Hz are the targets of several ground-based detectors like, LIGO \citep{ligo,ligo2}, VIRGO \citep{vir,virgo0,virgo}, TAMA \citep{Tsu,tama,tama2}, GEO \citep{geo2,geo}, etc. \textcolor{black}{Typical sources of such high-frequency waves are coalescing binary systems hosting stellar-sized compact objects like neutron stars and/or black holes; see, e.g., Sect. 6.1 of \citet{Flan05}.} The space-based LISA mission\footnote{{LISA as a mission first {appeared} in 1993 \citep{Danz} as a mission proposal to ESA's Horizon 2000 program.}} \citep{lisa0,lisa1,lisa}, now evolved into eLISA\footnote{See https://www.elisascience.org/ on the WEB.} \citep{elisa},  aimed to detect gravitational waves in the frequency range $10^{-5}-1$ Hz, while accurate timing measurements of pulsars may detect signals in the range $10^{-9}-10^{-7}$ Hz \citep{Kope,Hand,Jenet}. \textcolor{black}{Sect. 6.2 of \citet{Flan05} yields an overview of typical LISA sources: they include, e.g., equal mass binaries, in which the member black holes are of roughly equal mass ($\sim 10^5-10^8 M_{\odot}$), and extreme mass ratio binaries made by white dwarfs, neutron stars and $10-100 M_{\odot}$ black holes captured by $10^5-10^7 M_{\odot}$ black holes, located at quite large distances. As far as the very low frequency band is concerned ($10^{-9}-10^{-7}$ Hz), Sect. 6.3 of \citet{Flan05} deals with it. It may be due to several unresolved coalescing massive black holes; binaries which are either too massive to emit in the LISA band , or else are in-spiralling and will finally merge in several centuries or millennia.}

In this paper, we will analytically work out the long-term orbital variations of all the six standard Keplerian orbital elements of a solar system planet  due to the action of an externally incident monochromatic plane gravitational wave in the green-black\footnote{See Figure 1 of \citet{Prince}.} part of the spectrum, i.e. with frequency $\nu_g \ll 10^{-7}-10^{-10}$ Hz. Such kinds of gravitational waves  are important since they carry information about how  galaxies and black holes co-evolved over the history of the Universe \citep{Plow,Sesa},  the early universe and related exotic physical processes like, e.g., inflation and cosmic strings, and possible physics beyond the standard model of particles and fields \citep{Vilen,Ruba,Fabbri,Hogan86,Gris93}.
In particular, a primordial background of ultra-low frequency stochastic
gravitational waves with a characteristic spectral shape should be produced due to the coupling of the gravitational field with the exponential expansion driven by the inflation \citep{Gris75,Staro79,Ruba,Fabbri,Abbo84,Kra14}.
Concerning the possible existence of a background of gravitational waves dating back to the origin of the Universe, see, e.g., \citet{Web,Whel,Zel,Carr80}, and the discussion in \citet{Mash81}.

The effects of incident gravitational waves on the orbital motion of gravitationally bound systems were {inspected} by several authors with a variety of  approaches and approximations {pertaining} various features of both the waves and the orbits themselves \citep{Berto73,Rud75,Mash78,Futa79,Mash79,Tur79,Gris80,Mash81,Lin82a,Lin82b,Nel82,Iva87,Koc87,Chico96a,Chico96b,Isma11}. The idea of using the solar system to try to detect a stochastic background of gravitational waves of wavelengths much larger than about 1 au was first suggested by \citet{Berto73}.

{At first sight, the calculations presented here may be regarded just as an academic exercise with respect to empirical celestial mechanics, although such an allegation may sound  somewhat bizarre in view of the large amount of more or less analogous studies existing in literature concerning all sort of putative modified models of gravity based on much less solid theoretical and/or empirical support with respect to a GTR prediction like gravitational waves. Actually, it is not so for a variety of reasons. Indeed, recent developments in the field of planetary orbital determination, pursued by independent teams of astronomers \citep{Fie011,Pit13}, provided us with an increasing number of empirically estimated corrections $\Delta\dot\Ps$ to the standard Newtonian long-term precessions of some Keplerian orbital elements $\Ps$ for almost all the major bodies of the solar system. The times when the focus was solely on the perihelion of Mercury are definitively waned. Such corrections $\Delta\dot\Ps$ are, in general, determined in a purely phenomenological way, and may account for any unmodeled/mismodeled dynamical feature not included in the usual dynamical force models fit to the observations. The availability of increasingly extended data records of higher quality, and the forthcoming adoption of more accurate observational techniques \citep{Iess07,Iess09} will allow to reach unprecedented accuracies in knowing $\Delta\dot \Ps$.
This would allow to effectively put more and more stringent constraints on passing ultra-low frequency waves. This aspect is particularly important after the recent BICEP2 detection of B-mode polarization of the CMB at large angular scales \citep{BICEP1}. Indeed, it is of crucial importance to look for independent tests of the gravitational waves causing it. Moreover, the availability of exact analytical expressions for the wave-induced long-term orbital effects  may allow to set suitable linear combinations of corrections $\Delta\dot\Ps$ in order to separate them from unwanted, confusing orbital changes caused by other subtle standard physical effects like, e.g., solar and planetary oblateness, tides, minor bodies, etc. In this respect, it is also remarkable the fact that we have at our disposal empirically determined corrections $\Delta\dot\Ps$ for more than one planet. Moreover,} as far as the search for gravitational waves {in the solar system} is concerned, our results may be extended also to  spacecraft-based missions in the solar system like ASTROD-GW \citep{Ni,Men}, constituted of a number of probes in wide heliocentric orbits, and LISA \citep{Povo,Xia} envisaging the use of three spacecrafts
orbiting the Sun at 1 au distance in a quasi-equilateral
triangle formation 20 deg behind the Earth. {It has to be noticed that, in this cases, quite accurate devices to measure relative positions among the spacecraft would be used}. It is also the case of recalling that
the possibilities {offered} by Doppler tracking of interplanetary drag-free spacecraft to detect cosmological gravitational waves with wavelength of the order of, or larger than 1 au were studied in the past \citep{And71,MashGris,Bert}; for a recent review, see \citet{Arms} and references therein. Our calculations are valid, in principle, also for other natural systems like extrasolar planets\footnote{See http://exoplanet.eu/ on the WEB.} \citep{Torres} many of which have orbital frequencies of the order of $10^{-4}$ Hz, orbiting their parent stars at distances as small as $\simeq 0.01$ au. In such cases, our findings are technically valid for  waves with higher frequencies with respect {to} the solar system ones: indeed, they might be as high as about\footnote{However, for waves with such relatively high frequencies other facilities like LISA would be available: if they will finally become operative at the expected level of accuracy, they would likely surpass the possibilities offered by the extrasolar planets.} $10^{-6}$ Hz corresponding to the green-light blue part of the spectrum in Figure 1 of \citet{Prince}. {Another scenario to which our analysis can be applied is the stellar system orbiting the Supermassive Black Hole (SBH) hosted by the Galactic Center (GC) in Sgr A$^{\ast}$ \citep{Gille09}, where the orbital periods of the stars discovered so far are larger than 16 yr corresponding to frequencies smaller than $2\times 10^{-9}$ Hz.   }
On the other hand, our results are not necessarily limited to the very-low frequency waves case, being  valid for any tidal force with constant (over particle's characteristic timescales) matrix coefficients as well, independently of its physical origin.

The plan of the paper is as follows. In Section \ref{accz} we will discuss the analytical form of the acceleration experienced by the test particle due to an incident monochromatic plane gravitational wave traveling along a generic direction {$\bds{\hat{k}}$ of an arbitrary local Fermi frame}. We will also consider the simplified cases of a gravitational wave moving along the $z$ axis, as it is a choice widely adopted in literature, and in the reference $\{x,y\}$ plane. In Section \ref{keple} we will analytically work out the long-period changes occurring in the particle's orbital motion in the case of wave's frequencies much smaller than the orbital ones. We will  make {neither} any simplifying assumptions about the inclination and the eccentricity of the orbit {nor on the reference frame adopted}. In Section \ref{paragone} we briefly review some of the approaches followed in literature. Section \ref{conclu} is devoted to the conclusions.
\section{The acceleration imparted on an orbiting planet by a passing monochromatic plane gravitational wave}\lb{accz}
The action of an incoming  gravitational wave  on a planet of the solar system  is of tidal type with respect to a suitably constructed local inertial frame, represented by a Fermi coordinate system $\{x,y,z,t\}$,
whose origin is located at the Sun's position. In general, a tidal acceleration $\bds A$ experienced by a  slowly moving test particle due to an external curved spacetime metric can be written  in terms of the \virg{electric} components $R^i_{\ 0j0},\ i,j=1,2,3$ of  the local Riemann curvature tensor \textbf{\textsf{R}}
as {\citep{Misner}}
\eqi \ddot{x}^i\simeq -R^i_{\ 0j0}x^j\doteq \sum_{j=1}^3\mathcal{K}_{ij}x^j, i=,1,2,3,\lb{piar}\eqf
where we introduced the coefficients $\mathcal{K}_{ij}\doteq -R^i_{\ 0j0},\ i,j=1,2,3$ of the tidal matrix \textbf{\textsf{K}} {which are, dimensionally, $[\mathcal{K}_{ij}]={\rm T}^{-2}$}.

In the linearized weak-field and slow-motion approximation,
 and for the case of a propagating gravitational wave, one has
%
%
%
%
\eqi R^i_{\ 0j0}\simeq -\rp{1}{2}\rp{\partial^2 h_{ij}}{\partial t^2}, i,j=1,2,3.\lb{equaz}\eqf
%
%
%
%
%
In the case of a propagating plane wave, of the form\footnote{In \rfr{plana} $\chi^{\mu\nu},\ \mu,\nu=0,1,2,3$ represent the gravitational radiation field, {while $\mathbb{I}\doteq\sqrt{-1}$}.}
 \eqi h^{\mu\nu}={\rm Re}\left\{\chi^{\mu\nu}\exp\left[{\mathbb{I}}\left(\nu_g t-\bds{\hat{k}}\bds\cdot\bds r\right)\right]\right\},\ \mu,\nu=0,1,2,3,\lb{plana}\eqf
 %
 %
 %
 %
 %
 \textcolor{black}{as the linarized Rienmann tensor is gauge-invariant \citep{Misner}, one can compute $h_{00}$ and $h_{0i}, i=1,2,3$ directly in the transverse traceless\footnote{
 More general expressions for the TT metric tensor, not limited to the plane-wave and radiative approximation and with $h_{00}\neq 0,h_{0i}\neq 0, i=1,2,3$, can be found in \citet{Kop06,Kop011}; see also \citet{Poi07}. It can be seen that the parts not directly related to the wave are essentially identical to those used in the standard harmonic gauge used in ordinary planetary data reduction.} (TT) gauge \citep{Misner} finding that they vanish.}
\textcolor{black}{At this point, a clarification about a subtle issue pertaining the form of \rfr{piar} is in order. If, on the one hand, the tidal matrix elements ${\mathcal{K}}_{ij}$ are gauge-invariant, on the other hand, the matrix product yielding \rfr{piar} is gauge-dependent. In GTR, a gauge transformation brings with it a coordinate transformation. In general, the Fermi coordinates are different from the TT coordinates. Suffice it to say that, in the TT gauge, it is $h_{00} = 0$, thus eliminating the Newtonian potential and making impossible any description of orbital mechanics because of lacking of the central body. Nonetheless, as a general remark by \citet{bask04}, even if one starts  from the general form of the wave's field, which includes also the non-TT components, one would end up with equations of motion containing only the TT-components. It is so because the equation for the geodesic deviation involves the curvature tensor (and its derivatives) in which the non-TT components automatically cancel out. Moreover, the transformation between the Fermi and the TT coordinates \citep{bask04} is proportional to the ratio of the binary's spatial extension to the wavelength and to its time derivative, which in our case are negligible. This would allow to neglect also the \virg{magnetic}-type components entering the equations of motion in addition to the \virg{electric} ones yielding just \rfr{piar}.}

The tidal matrix \textbf{\textsf{K}} is not only symmetric, but also traceless. It has five independent components, so that the acceleration $\bds A$ of \rfr{piar} felt by a test particle becomes, quite generally,
\begin{equation}
\left\{
\begin{array}{lll}
A_x & = & \mathcal{K}_{xx}x + \mathcal{K}_{xy}y + \mathcal{K}_{xz}z, \\ \\
A_y & = & \mathcal{K}_{xy}x + \mathcal{K}_{yy}y + \mathcal{K}_{yz}z, \\ \\
A_z & = & \mathcal{K}_{xz}x + \mathcal{K}_{yz} y-(\mathcal{K}_{xx}+\mathcal{K}_{yy})z.
\end{array}
\right.\lb{maronna}
\end{equation}
In fact, the condition that the force  exerted on the  particle    by the wave  is orthogonal to its direction of propagation
yields three further  constraints \begin{equation}
\left\{
\begin{array}{lll}
\mathcal{K}_{xx}\hat{k}_x + \mathcal{K}_{xy}\hat{k}_y +\mathcal{K}_{xz}\hat{k}_z & = & 0, \\ \\
\mathcal{K}_{xy}\hat{k}_x + \mathcal{K}_{yy}\hat{k}_y +\mathcal{K}_{yz}\hat{k}_z & = & 0, \\ \\
\mathcal{K}_{xz}\hat{k}_x + \mathcal{K}_{yz}\hat{k}_y -\left(\mathcal{K}_{xx}+\mathcal{K}_{yy}\right)\hat{k}_z & = & 0,
\end{array}
\right.\lb{condiz}
\end{equation}
so that the independent components of $\mathcal{K}_{ij},\ i,j=1,2,3$ are just two. In the case of \rfr{plana}, they are harmonic functions proportional to $\nu^2_g$.
%
%

For a wave not traveling in the reference $\{x,y\}$ plane, i.e. for $\hat{k}_z\neq 0$, \rfr{condiz} yields
\begin{equation}
\left\{
\begin{array}{lll}
\mathcal{K}_{xz} & = &-\mathcal{K}_{xx}\left(\rp{\hat{k}_x}{\hat{k}_z}\right)-\mathcal{K}_{xy}\left(\rp{\hat{k}_y}{\hat{k}_z}\right), \\ \\
\mathcal{K}_{yy}& = &-\mathcal{K}_{xx}\left(\rp{\hat{k}_x^2+\hat{k}_z^2}{\hat{k}_y^2+\hat{k}_z^2}\right)-2\mathcal{K}_{xy}\left(\rp{\hat{k}_x \hat{k}_y}{\hat{k}_y^2+\hat{k}_z^2}\right), \\ \\
\mathcal{K}_{yz} & = &\mathcal{K}_{xx}\left[\rp{\hat{k}_y}{\hat{k}_z}\left(\rp{\hat{k}_x^2+\hat{k}_z^2}{\hat{k}_y^2+\hat{k}_z^2}\right)\right]+
\mathcal{K}_{xy}\left\{\left(\rp{\hat{k}_x}{\hat{k}_z}\right)\left[\rp{2\hat{k}_y^2}{\left(\hat{k}_y^2+\hat{k}_z^2\right)}-1\right]\right\}.
\end{array}\lb{zimba}
\right.
\end{equation}
Thus,  we pose
\begin{equation}
\left\{
\begin{array}{lll}
h_1 & \doteq & \mathcal{K}_{xx}, \\ \\
h_2 & \doteq & \mathcal{K}_{xy}
\end{array}
\right.\lb{h1h2}
\end{equation}
for the two independent polarizations of the gravitational wave.

When the wave propagates in the reference $\{x,y\}$ plane, i.e. for $\hat{k}_z=0$, the second equation in \rfr{zimba} becomes
\eqi \mathcal{K}_{yy}=-\mathcal{K}_{xx}\left(\rp{\hat{k}_x}{\hat{k}_y}\right)^2 -2\mathcal{K}_{xy}\left(\rp{\hat{k}_x}{\hat{k}_y}\right),\lb{pian} \eqf
while from \rfr{condiz} it turns out
\begin{equation}
\left\{
\begin{array}{lll}
\mathcal{K}_{xy} & = &-\mathcal{K}_{xx}\left(\rp{\hat{k}_x}{\hat{k}_y}\right), \\ \\
\mathcal{K}_{yz} & = &-\mathcal{K}_{xz}\left(\rp{\hat{k}_x}{\hat{k}_y}\right).
\end{array}\lb{iuy}
\right.
\end{equation}
Thus, in this case, we define
\begin{equation}
\left\{
\begin{array}{lll}
h_1 & \doteq & \mathcal{K}_{xx}, \\ \\
h_2 & \doteq & \mathcal{K}_{xz}.
\end{array}
\right.\lb{h1h2j}
\end{equation}
Notice that both \rfr{pian} and \rfr{iuy} do not hold for a wave traveling along the reference $x$ axis, i.e. for $\hat{k}_y=0$.

Finally, let us notice that when $\bds{\hat{k}}=\{\pm 1,0,0\}$, \rfr{condiz} tells us that
\eqi \mathcal{K}_{xx}=\mathcal{K}_{xy}=\mathcal{K}_{xz}=0,\lb{etnow}\eqf so that
it can be posed
\begin{equation}
\left\{
\begin{array}{lll}
h_1 & \doteq & \mathcal{K}_{yy}, \\ \\
h_2 & \doteq & \mathcal{K}_{yz}.
\end{array}
\right.\lb{h1h2new}
\end{equation}

In order to make contact with realistic situations occurring in typical solar system data analyses, we remark that the unit vector $\bds{\hat{k}}$ can be written, in general, as
\begin{equation}
\left\{
\begin{array}{lll}
\hat{k}_x & = & \cos\beta\cos\lambda, \\ \\
\hat{k}_y & = &\cos\beta\sin\lambda, \\ \\
\hat{k}_z & = & \sin\beta
\end{array}\lb{rbetl}
\right.
\end{equation}
in terms of the ecliptic latitude $\beta$ and longitude $\lambda$, which are to be {considered} as unknown{: in this case, the reference $\{x,y\}$ plane would typically coincide with the mean ecliptic at the epoch J$2000.0$.}
Explicit expressions of the tidal matrix coefficients for a generic wave's incidence can be found in \citet{Chico96a}. In addition to the  amplitudes of the two independent wave's polarizations and of their mutual constant phase difference, they contains $\beta$ and $\lambda$ through the angle $\Theta$ \citep{Chico96a}. It is defined from
$\bds{\hat{k}}\bds\cdot\bds{\hat{N}}=\cos\Theta$ and $\left|\bds{\hat{k}}\bds\times\bds{\hat{N}}\right|=\sin\Theta$, where $\bds{\hat{N}}$ is the unit vector directed along the test particle's out-of-plane direction
coinciding with the direction of the orbital angular momentum.

Inserting \rfr{zimba}, with \rfr{rbetl}, in \rfr{maronna} yields\footnote{Notice that \rfr{strazza} has a singularity for $\beta=0$, i.e. for a wave traveling in the reference $\{x,y\}$ plane.}
\begin{equation}
\left\{
\begin{array}{lll}
A_x & = & h_{1}\left(x-z\cot\beta\cos\lambda\right)+h_{2}\left(y-z\cot\beta\sin\lambda\right), \\ \\
A_y & = & h_{1}\left\{-y\left(\rp{\sin^2\beta+\cos^2\beta\cos^2\lambda}{\sin^2\beta+\cos^2\beta\sin^2\lambda}\right)+
z\left[\rp{\cot\beta\sin\lambda\left(\sin^2\beta+\cos^2\beta\cos^2\lambda\right)}{\sin^2\beta+\cos^2\beta\sin^2\lambda}\right]\right\}+ \\ \\
&+& h_{2}\left[x-y\left(\rp{\cos^2\beta\sin 2\lambda}{\sin^2\beta+\cos^2\beta\sin^2\lambda}\right)-
z\cot\beta\cos\lambda\left(\rp{\sin^2\beta-\cos^2\beta\sin^2\lambda}{\sin^2\beta+\cos^2\beta\sin^2\lambda}\right)\right], \\ \\
A_z & = & h_{1}\left\{-x\cot\beta\cos\lambda+y\left[\rp{\cot\beta\sin\lambda\left(\sin^2\beta+\cos^2\beta\cos^2\lambda\right)}{\sin^2\beta+\cos^2\beta\sin^2\lambda}\right]
+z\left(\rp{\cos^2\beta\cos 2\lambda}{\sin^2\beta+\cos^2\beta\sin^2\lambda}\right)\right\}+\\ \\
&+& h_{2}\left[-x\cot\beta\sin\lambda-y\cot\beta\cos\lambda\left(\rp{\sin^2\beta-\cos^2\beta\sin^2\lambda}{\sin^2\beta+\cos^2\beta\sin^2\lambda}\right)
+z\left(\rp{\cos^2\beta\sin 2\lambda}{\sin^2\beta+\cos^2\beta\sin^2\lambda}\right)\right].
\end{array}\lb{strazza}
\right.
\end{equation}
%
%
%
In the specific case of a plane wave propagating along the reference $z$ axis, i.e. for $\beta=\pm 90$ deg,
%
%
\rfr{condiz} and \rfr{zimba} yield
\begin{equation}
\left\{
\begin{array}{lll}
\mathcal{K}_{yy}& = &-\mathcal{K}_{xx}, \\ \\
\mathcal{K}_{xz}& = & 0, \\ \\
\mathcal{K}_{yz}& = & 0.
\end{array}\lb{prenci}
\right.
\end{equation}
Thus, \rfr{strazza} {reduces} to the well known result \citep{Nel82,Iva87,Isma11}
\begin{equation}
\left\{
\begin{array}{lll}
A_x & = & h_1 x+ h_2 y, \\ \\
A_y & = & h_2 x-h_1 y, \\ \\
A_z & = & 0.
\end{array}
\right.\lb{agw}
\end{equation}

For $\beta=0$, i.e. for $\hat{k}_z=0$,  \rfr{maronna} {reduces} to\footnote{Notice that \rfr{nueva} has a singularity for $\lambda=0$, i.e. for a wave traveling along the reference $x$ axis.}
\begin{equation}
\left\{
\begin{array}{lll}
A_x & = & h_1 \left(x-y\cot\lambda\right)+ h_2 z, \\ \\
A_y & = & -\cot\lambda\left[h_1\left(x-y\cot\lambda\right)+h_2 z\right], \\ \\
A_z & = & h_1 z \left(\rp{\cos 2\lambda}{\sin^2\lambda}\right)+h_2\left(x-y\cot\lambda\right)\lb{nueva}
\end{array}
\right.
\end{equation}
because of \rfr{pian}, \rfr{iuy} and \rfr{rbetl}. Notice that \rfr{nueva} is not singular for  $\bds{\hat{k}}=\{0,\pm 1,0\}$, and $A_x\neq 0,A_y=0,A_z\neq 0$, as it is expected for a plane wave.

Finally, for a wave traveling along the reference $x$ axis \rfr{maronna}, with \rfr{etnow} and \rfr{h1h2new}, becomes
\begin{equation}
\left\{
\begin{array}{lll}
A_x & = & 0, \\ \\
A_y & = & h_1 y + h_2 z, \\ \\
A_z & = & h_2 y -h_1 z.\lb{nueva2}
\end{array}
\right.
\end{equation}

In Section \ref{keple} we will analytically work out the effects of \rfr{maronna}  on the trajectory of a test particle orbiting a central body with gravitational parameter $GM$, where $G$ is the Newtonian gravitational constant and $M$ is its mass, located at the origin of the Fermi frame traversed by a  monochromatic plane gravitational wave. We will also consider the particular case of of \rfr{agw} ($\bds{\hat{k}}=\{0,0,\pm 1\}$), widely treated in literature.
\section{The long-term variations of the Keplerian orbital elements}\lb{keple}
The typical planetary orbital frequencies $\nk$
in the solar system vary from $\nk = 1.3\times 10^{-7}$ Hz (Mercury) to $\nk = 1.2\times 10^{-10}$ Hz (Pluto), corresponding to timescales $P_{\rm b}$
ranging from $7\times 10^6$ s (Mercury) to $8\times 10^9$ s (Pluto). They are much larger than the time {required} by the light to travel across the spatial extensions of the Sun's planetary orbits, ranging from $2\times 10^2$ s (Mercury) to $2\times 10^4$ s (Pluto). Thus, if we consider the action of a monochromatic plane gravitational wave of frequency $\nu_g$ and wave vector $\bds k$, with $k=\nu_g/c$, on a planetary orbit during a time interval $\Delta t$ comparable to an orbital period $P_{\rm b}$, its phase $\Phi\doteq\nu_g t-\bds k\cdot \bds r=\nu_g[t-(r/c)\cos \alpha]$ can be reasonably approximated by $\Phi\simeq \nu_g t$, independently of the orientation $\alpha$ of $\bds k$ with respect to the planet's position $\bds r$. As previously stated, in the rest of this Section we will  assume $\nu_g/\nk\ll 1$ as well.

\subsection{Monochromatic plane gravitational wave propagating along a generic direction}
{A} straightforward {first-order perturbative} calculation {performed with the standard Gauss equations for the variations of the \textcolor{black}{osculating Keplerian orbital} elements \citep{BeFa}}
%
%
yield{s} the long-term, i.e. averaged over one orbital period, variations of all the osculating Keplerian orbital elements of the test particle due to \rfr{maronna}.
\textcolor{black}{We briefly recall the usual computational procedure.  A standard Keplerian ellipse is assumed as reference, unperturbed orbit; any small additional acceleration like, e.g., \rfr{maronna} is considered as a perturbation; the Gauss equations are valid for any kind of perturbation, irrespectively of its physical origin. The disturbing acceleration is inserted into the right-hand-sides of the Gauss equations, which are evaluated onto the unperturbed orbit. Then, an average over one full orbital revolution of the test particle is performed  to obtain the long-term orbital variations. We notice that the analytical expression of \rfr{maronna} was obtained in the TT gauge, which is a particular case of the harmonic gauge: from the point of view of the pertubative calculation using the Newtonian two-body problem as zeroth order term, it does not pose problems since the latter one  is gauge-invariant \citep{Flan05}. It clearly appears also in the explicit expression for, say, $h_{00}$ in the TT gauge when a central mass is present in the local frame traversed by the external wave \citep{Kop06,Kop011,Poi07}. In principle, one could also take  different reference orbits already including 1PN Schwarzschild-like terms \citep{calura1,calura2}; in such a case additional mixed, Schwarzschild-radiative terms would arise, but they would be negligible because of higher order in powers of $c^{-1}$. As a final remark to  adequately put the problem in context, we recall that we are in a linearized, weak-field and slow-motion scenario. We are not dealing here with the final merging stage of a two-body system made by highly relativistic, self-gravitating compact objects whose orbits are shrinking because of the emission of gravitational waves from the system itself.}

\textcolor{black}{The long-term orbital variations due to \rfr{maronna} are, thus}
{\tiny{\begin{equation}
\left\{
\begin{array}{lll}
\dert a t & = & 0, \\ \\
\dert e t & = & \rp{5e\sqrt{1-e^2}}{16 \nk}\left\{
-8\sin I\cos 2\omega\left(\kxz\cos\Om+\kyz\sin\Om\right)+\right.\\ \\
&+&\left. 4\cos I\cos 2\omega\left[-2\kxy\cos 2\Om +\left(\kxx-\kyy\right)\sin 2\Om\right]+\right.\\ \\
&+&\left.\sin 2\omega\left[
\left(\kxx-\kyy\right)\left(3+\cos 2 I\right)\cos 2\Om + 6\left(\kxx+\kyy\right)\sin^2 I+\right.\right. \\ \\
&+&\left.\left. 4\sin 2 I\left(\kxz\sin\Om-\kyz\cos\Om\right)+2\kxy\left(3+\cos 2 I\right)\sin 2\Om
\right]\right\},\\ \\
\dert I t &=& \rp{1}{16\nk\sqrt{1-e^2}}\left[|
4\cos I\left(2+3e^2+5e^2\cos 2\omega\right)\left(\kxz\cos\Om+\kyz\sin\Om\right)-\right. \\ \\
&-&\left. 2\left(2+3e^2\right)\sin I\left[2\kxy\cos 2\Om +\left(\kyy-\kxx\right)\sin 2\Om\right]-\right. \\ \\
&-&\left.10 e^2\sin I\cos 2\omega\left[2\kxy\cos 2\Om +\left(\kyy-\kxx\right)\sin 2\Om\right]+\right. \\ \\
&+&\left. 5e^2\sin2\omega\left\{4\cos 2 I\left(\kyz\cos\Om-\kxz\sin\Om\right)+\right.\right.\\ \\
&+&\left.\left.\sin 2 I\left[-3\left(\kxx+\kyy\right)+\left(\kxx-\kyy\right)\cos 2\Om+2\kxy\sin 2\Om\right]\right\}
 | \right], \\ \\
 \dert\Om t & =& -\rp{1}{8\nk\sqrt{1-e^2}}\left\{ 2\cos 2I\csc I\left(-2-3e^2+5e^2\cos 2\omega\right)\left(\kyz\cos\Om-\kxz\sin\Om\right)+\right. \\ \\
 &+&\left.\cos I\left(-2-3e^2+5e^2\cos 2\omega\right)\left[-3\left(\kxx+\kyy\right)+\left(\kxx-\kyy\right)\cos 2\Om+2\kxy\sin2\Om\right]+\right.\\ \\
 &+&\left. 5e^2\sin 2\omega\left[2\kxy\cos 2\Om-2\cot I\left(\kxz\cos\Om+\kyz\sin\Om\right)+\left(\kyy-\kxx\right)\sin 2\Om\right]
 \right\}, \\ \\
 \dert\varpi t &=& \rp{1}{16\nk\sqrt{1-e^2}}\left\{
 -3\left(\kxx+\kyy\right)\left(1-4\cos I+\cos 2 I\right)\left(-2-3e^2+5e^2\cos 2\omega\right)+\right. \\ \\
 &+&\left.\left(\cos I-2\right)\cos 2\Om\left[2\left(\kxx-\kyy\right)\cos I\left(-2-3e^2+5e^2\cos 2\omega\right)+20e^2\kxy\sin 2\omega\right]+\right. \\ \\
 &+&\left.2\left(\cos I-2\right)\cos\Om\csc I\left[2\kyz \cos 2I\left(-2-3e^2+5e^2\cos 2\omega\right)-10e^2\kxz\cos I\sin 2\omega\right]+\right. \\ \\
 &+&\left.2\left(\cos I-2\right)\sin\Om\csc I\left[-2\kxz \cos 2I\left(-2-3e^2+5e^2\cos 2\omega\right)-10e^2\kyz\cos I\sin 2\omega\right]+\right. \\ \\
 &+&\left.2\left(\cos I-2\right)\sin 2\Om\left[2\kxy \cos I\left(-2-3e^2+5e^2\cos 2\omega\right)+5e^2\left(\kyy-\kxx\right)\sin 2\omega\right]
 \right\}, \\ \\
 \dert{\mathcal{M}}t  & = \nk + &\rp{1}{32\nk}\left\{
 8\left[-7-3e^2+5\left(1+e^2\right)\cos 2\omega\right]\sin 2 I\left(\kyz\cos\Om-\kxz\sin\Om\right)-\right.\\ \\
 &-&\left. 80\left(1+e^2\right)\sin I\sin 2\omega\left(\kxz\cos\Om+\kyz\sin\Om\right)+\right. \\ \\
 &+&\left. 2\cos 2 I\left[-7-3e^2+5\left(1+e^2\right)\cos 2\omega\right]\left[3\left(\kxx + \kyy\right)+\left(\kyy-\kxx\right)\cos 2\Om-2\kxy\sin 2\Om\right]-\right. \\ \\
 &-& 2\left.\left[7+3e^2+5\left(1+e^2\right)\cos 2\omega\right]\left[\kxx+\kyy+\left(\kxx-\kyy\right)\cos 2\Om + 2\kxy\sin 2\Om\right]-\right.\\ \\
 &-&\left.40\left(1+e^2\right)\cos I\sin 2\omega\left[2\kxy\cos 2\Om+\left(\kyy-\kxx\right)\sin 2\Om\right]
 \right\}.
\end{array}\lb{superrates}
\right.
\end{equation}}
}
In order to deal with manageable expressions of general validity, we did not display in \rfr{superrates} the explicit expressions of the tidal matrix coefficients in terms of {any specific representation of $\bds{\hat{k}}$. We stress that  \rfr{superrates} is valid for a generic reference frame. Moreover, it}
  is exact both in the {eccentricity} $e$ and in the {inclination} $I$ {to the reference $\{xy\}$ plane} in the sense that no \textit{a priori} simplifying assumptions about the eccentricity and the inclination of the perturbed test particle's orbit  were assumed in the calculation. It can be noticed that the semi-major axis {$a$} is not affected by the passage of a very slowly varying gravitational wave; the variation of the eccentricity is proportional to $e$ itself, so that a circular orbit does not change its shape. {It is important to remark that, since in the real world exactly circular orbits do not exist, the previous result do not mean that the shape of the orbit is left unaffected by a passing ultra-low plane gravitational wave. Indeed, as previously stated, $e$ characterizes just the orbit's appearance, and its change in time implies that, e.g., the aphelion and perihelion distances $r_{\rm max}=a(1+e),r_{\rm min}=a(1-e)$ vary even if $a$, which is responsible of the orbit's size, does not. In other words,  the ellipse would remain\footnote{{Actually, also its orientation within its plane itself would change because of the variation of $\varpi$.}} inscribed within the same imaginary circle, but it would get more or less elongated as time passes.  } Moreover, \rfr{superrates} does not contain {genuine} secular effects because of the presence of\footnote{{Here $\omega$ denotes the argument of pericenter, while $\varpi\doteq \Om+\omega$ is the argument of latitude.}} $I,\Om,\omega$ which, actually, experience slow changes\footnote{They are mainly due to the mutual N$-$body classical perturbations: the rates of change can be found at http://ssd.jpl.nasa.gov/txt/p$\_$elem$\_$t1.txt.} in time as far as the planets of the solar system are concerned. A further modulation is due to  the coefficients $\mathcal{K}_{ij}$ of the tidal matrix containing the (ultra-low) harmonic variation of the impinging gravitational wave. Since all such frequencies are much smaller than the typical orbital ones for the Sun's planet, the terms containing them were kept fixed in the integration yielding \rfr{superrates}. {Finally, we notice that the temporal signatures of all the long-term variations of \rfr{superrates} are peculiar to the action of just\footnote{This is true if the standard Newtonian-Einsteinian laws of gravitation are involved.} an incident ultra-low plane gravitational wave. This would be of great help in recognizing its passage if and when orbital variations matching the specific patterns of \rfr{superrates}  will be, actually, detected over multidecadal analyses.}
\subsection{Monochromatic plane gravitational wave propagating along  the $z$ axis}
By using the same procedure it is possible to work out the long-term variations of the Keplerian orbital elements for a   direction of incidence of the wave coinciding with the reference $z$ axis. By using \rfr{agw} one gets
\begin{equation}
\left\{
\begin{array}{lll}
\dert a t & = & 0, \\ \\
\dert e t & = & \rp{5e\sqrt{1-e^2}\left[-8\cos I\cos 2\omega\left(h_2\cos 2\Om-h_1\sin 2\Om\right)+2\left(3+\cos 2 I\right)\sin 2\omega\left(h_1\cos 2\Om+h_2\sin 2\Om\right)\right]}{16 \nk}, \\ \\
\dert I t & = & \rp{\sin I \left[-\left(2+3e^2+5e^2\cos 2\omega\right)\left(h_2\cos 2\Om-h_1\sin 2\Om\right)+5e^2\cos I\sin 2\omega\left(h_1\cos 2\Om+h_2\sin 2\Om\right)\right]}{4\nk\sqrt{1-e^2}}, \\ \\
\dert \Om t &=& -\rp{5e^2\sin 2\omega\left(h_2\cos 2\Om-h_1\sin 2\Om\right)+\cos I\left(-2-3 e^2+5 e^2\cos 2\omega\right)\left(h_1\cos 2\Om+h_2\sin 2\Om\right)}{4\nk\sqrt{1-e^2}},\\ \\
\dert\varpi t &=& \rp{1}{32 \nk\sqrt{1-e^2}}\left\{ -40\left[e^2+\left(e^2-2\right)\cos I\right]\sin 2\omega\left(h_2\cos 2\Om-h_1\sin 2\Om\right)+\right., \\ \\
 & + & \left. 20\left(3-2e^2-2 e^2\cos I+\cos 2 I\right)\cos 2\omega\left(h_1\cos 2\Om+h_2\sin 2\Om\right)+\right. \\ \\
 &+&\left. 16\left(3-3e^2+5\cos I\right)\sin^2\left(\rp{I}{2}\right)\left(h_1\cos 2\Om+h_2\sin 2\Om\right)\right\}, \\ \\
 \dert{\mathcal{M}} t  & = & \nk + \rp{1}{32 \nk}\left[-80\left(1+e^2\right)\cos I\sin2\omega\left(h_2\cos 2\Om-h_1\sin 2\Om\right)-\right. \\ \\
 &-& \left. 20\left(1+e^2\right) \left(3+\cos 2 I\right)\cos 2\omega\left(h_1\cos 2\Om+h_2\sin 2\Om\right)-\right. \\ \\
 &-& \left. 8\left(7+3e^2\right)\sin^2 I\left(h_1\cos 2\Om+h_2\sin 2\Om\right)\right].
\end{array}\lb{rates}
\right.
\end{equation}
Also in this case, no a priori assumptions on $e$ and $I$ were made, so that the rates of \rfr{rates} are exact in this respect. Notice that for $I=0$, i.e. for incidence of the  gravitational wave normal to the orbital plane, the inclination is left unaffected. Moreover, also in this case, the elements $h_1$ and $h_2$ of the tidal matrix, which are (slowly) time-varying harmonic functions whose amplitudes are proportional to $\nu^2_g$, were assumed constant in the integrations over the planet's orbital period.

Notice that, for a plane wave traveling along the $z$ axis, i.e. for \rfr{prenci}, \rfr{superrates}  {reduces} just to \rfr{rates}. It is straightforward to obtain the formulas valid also for the other specific directions of propagation  examined in Section \ref{accz} by suitably specializing \rfr{superrates} to such cases (cfr. \rfr{pian}, \rfr{iuy} and \rfr{etnow}).
\subsection{A comment on the approximation used for the harmonic wave functions}
In obtaining \rfr{superrates} and \rfr{rates} we kept the tidal matrix coefficients, which include the time-dependent harmonic functions of the wave, constant in the integrations over one orbital revolution of the test particle. This implies that we {considered} waves having frequencies $\nu_g$ much smaller than the orbital ones $n$. As we will see  in Section \ref{paragone}, it is in contrast with the other approaches followed so far in literature, in which no similar assumptions  on $\nu_g$ were made. Apart from the fact that, from a computational point of view, our approach allows for exact calculations in $e$ and $I$, our choice is justified also from a phenomenological point of view. Indeed, applying our results to the solar system implies that we could, in principle, put constraints over a part of the ultra-low frequency spectrum of gravitational waves for which neither  ground-based nor space-based dedicated experimental devices are  available\footnote{A possible exception may be the proposed Inflation Probe by NASA \citep{Infla1}, dedicated to map the polarization of the Cosmic Microwave Background at $10^{-16}$ Hz (see Figure 1 of \citet{Prince}). Notice also that precision timing of millisecond pulsars may be used for gravitational waves in the range $10^{-7}-10^{-9}$ Hz. \citep{Kope,Hand,Jenet}.}. Moreover, in view of  continuous tracking of planets by means of  ranging either directly to their surfaces or to ever more numerous orbiting spacecraft  it will be possible, in principle, to obtain more and more accurate bounds because of increasing data records over the years. On the other hand, it is, after all, of little practical utility to perform cumbersome calculations of the wave-induced orbital effects involving frequencies as large as, or even larger than, the planet's ones since much more accurate dedicated experiments already exist covering such  windows of the spectrum of gravitational waves. Indeed, the accuracy reachable  in future planned interplanetary laser ranging facilities \citep{plr1,plr2,plr3}, of the order of about $1-10$ cm, is not comparable with that of the latest gravitational wave detectors; it can be acceptable in order to put upper bounds when no other, more accurate devices exist, as for the ultra-low frequency waves.
\section{Confrontation with other approaches in literature}\lb{paragone}
In this Section we briefly outline the approaches followed by some other researchers, with particular care to the orbital effects caused by the traveling wave.

\citet{Rud75} uses polar coordinates in the orbital plane and considers an orbit disturbed  by a  monochromatic plane gravitational wave traveling along a generic direction; the additional force resulting on the test particle is obtained within the linear approximation of general relativity (within the framework of the \virg{theory of gravitation in plane space} by \citet{Zeld}). Then, \citet{Rud75} assumes a normally incident  wave and solves the equations of motion at zero order in eccentricity. Finally, he studies the variation $\Delta r$ of the orbital radius  for various values of the wave's frequency.

\citet{Tur79} works in the TT gauge by deriving the geodesic equations of motion in cartesian coordinates of both the binary system's constituents. Then, he takes their difference\textcolor{black}{s} obtaining the equations for their relative motion. It turns out that they are formally different from those obtained by other authors like, e.g., \citet{Mash78} from the geodesic deviation equation, but \citet{Tur79} shows that they are, actually, equal up to an unphysical coordinate transformation. In the scenario by \citet{Tur79}, the monochromatic plane wave travels along the negative $z$ direction.  He uses the Gaussian perturbative scheme to work out the temporal changes over one orbital revolution of the semi-major axis and the eccentricity of a circular orbit by setting $\Om=\varpi=I=0$. Then, he discards the limitation $I=0$, but not the other ones. Finally, \citet{Tur79} considers non-circular orbits, and computes the variation of $a$ to the lowest order in $e$ for $\Om=I=\varpi=0$. In all of such cases, he takes different values for $\nu_g/\nk$.

\citet{Mash81}, relying upon the methods developed in \citet{Mash78}, look at the effects that various kinds of gravitational radiation, among which monochromatic plane waves are {considered} as well, have on a Newtonian binary system. Let us recall that \citet{Mash78}, in considering slow particle motions and weak waves having wavelengths larger than the system's orbital size, adopts the linearized Jacobi equation for the geodesic deviation equation. \citet{Mash78} writes down the equations of motion in cylindrical coordinates and solve them for various values of $\nu_g$ with respect to $\nk$. As far as the monochromatic plane wave case is concerned, also \citet{Mash81} assume that its wavelength is much larger than the size of the binary system; no simplifying assumptions on the wave's frequency are made. Then, \citet{Mash81} work out the resulting relative change $\Delta r/r$ occurring in the test particle's distance from the primary in the low-eccentricity approximation. Moreover, \citet{Mash81} consider also the case of a pair of planets moving in the same plane along circular orbits, and calculate the wave-induced relative change $\Delta\mathcal{R}/\mathcal{R}$ in their mutual distance.

\citet{Nel82} look in the TT gauge at a monochromatic plane gravitational wave traveling along the $z$ axis, and having wavelength much larger than the size of the system {considered}. They write down the components of the wave-induced acceleration experienced by the test particle in cylindrical coordinates for generic shape and inclination of the orbit. Then, they specialize them to the case of normally incident wave, i.e. for an orbit with zero inclination. At this point, \citet{Nel82} solve the resulting perturbed equations of motion by evaluating the radial change $\Delta r$ over one orbital revolution. In their computation, they resort to some approximations in $e$, and do not consider the harmonic wave functions constant over typical orbital timescales. The resulting changes for different values of $\nu_g/\nk$ are inspected. Then,  \citet{Nel82} examine the case of a generic inclination of the test particle's orbit by working out the shift along the wave's propagation, i.e. along the $z$ axis. Finally, \citet{Nel82} perform some numerical analyses of the radial shifts for different values of the eccentricity and of the phases of the wave.

Also \citet{Iva87} considers the case of a monochromatic plane gravitational wave, in the TT gauge, traveling along the $z$ axis. He assumes that its wavelength is much larger than the characteristic size of the perturbed system, so that he can neglect the term $\bds k\bds\cdot\bds r$ in the wave's phase. On the other hand, \citet{Iva87} does not make any a priori assumptions about the magnitude of the wave's frequency $\nu_g$ with respect to the orbital one $\nk$. After having obtained the exact radial, transverse and normal components of the perturbing acceleration from its expression in cartesian coordinates, \citet{Iva87} makes use of the Gauss equations for the variation of all the Keplerian orbital elements, apart from the mean anomaly. At this point, after having set $\omega=0$, he makes an expansion of the right-hand sides of the Gauss equations, evaluated onto the unperturbed Keplerian ellipse, to first order in $e$ by using the mean anomaly $\mathcal{M}$ as independent variable. Thus,
\citet{Iva87} obtains approximate expressions of the form \eqi \dert\xi t = \mathfrak{F}(t),\ \xi=a,e,I,\Om,\omega,\eqf which he straightforwardly integrates. In doing that, he does not consider $h_1$ and $h_2$ as constant by keeping, instead, their harmonic time dependence. Then, \citet{Iva87} discusses various particular cases for different orbital geometries and, especially, for different values of $\nu_g/\nk$.

Also \citet{Koc87} adopt a wave in TT gauge traveling along the reference $z$ axis. Moreover, they assume for it a circular polarization, and do not make any assumptions about the wave's frequency and its wavelength. \citet{Koc87} use cylindrical coordinates to write down  the spacetime line element for a linearized superposition of the Schwarzschild and wave fields, from which the particle's equations of motion are obtained.  Concerning the orbital geometry, they put  the perturbed orbit, not assumed a priori circular,  in the reference $\{x,y\}$ plane, so that $\bds k\bds\cdot\bds r=0$. Then, \citet{Koc87}  do not recur to any known perturbative schemes, like, e.g., the Gauss and the Lagrange equations \citep{BeFa}, in dealing with the resulting equations of motion. After some changes of variables, they integrate the resulting modified equations over one orbital revolution by inferring  the cumulative changes of the semi-major axis, the eccentricity and the pericenter after one orbital period. Finally, \citet{Koc87} consider how the results they obtained vary for different values of $\nu_g/\nk$; they consider the circular case as well.

\citet{Chico96a} and \citet{Chico96b} look at the action of a monochromatic plane gravitational wave traveling along the reference $z$ axis and normally impinging on an unperturbed two-body system. The wave's wavelength is assumed to be much larger than the semi-major axis of the orbit, and the characteristic velocities within the system are much smaller than $c$. No assumptions are made a priori on the wave's frequency $\nu_g$. The resulting relative acceleration is obtained from the geodesic deviation equation in cartesian coordinates. Then,
\citet{Chico96a} and \citet{Chico96b} write down the Hamiltonian of the perturbed system in polar coordinates, and adopt the Delaunay orbital elements $L,G,\ell,g$ for a non-circular orbit. Fourier series expansions in terms of the mean anomaly are performed. Different values for $\nu_g/\nk$ and wave's polarizations are, then, {considered}.

\citet{Jaco}, after having generalized the Jacobi equation by taking into account also the relative velocity of the geodesics, consider a monochromatic plane gravitational wave traveling along the reference $x$ axis of the local Fermi quasi-inertial frame. \citet{Jaco} write down the resulting equations of motion of the test particle in cartesian coordinates; they are rather involved because of the presence of the non-linear, velocity-dependent terms. Then, \citet{Jaco} study the motion of the test particle along the $x$ axis itself.

The general wave-binary system scenario adopted by \citet{Isma11} is analogous to that by, e.g., \citet{Iva87}. However, after having written the components of the tidal-type wave acceleration in cartesian coordinates, \citet{Isma11}  obtain a potential function $\mathcal{R}$ from them in order to use the perturbative scheme based on the Lagrange planetary equations \citep{BeFa} for all the Keplerian orbital elements, apart from $\mathcal{M}$. Then, they evaluate $\mathcal{R}$ onto the unperturbed Keplerian ellipse, without fixing $\omega$, and expand it to some, unspecified, order in $e$ by using the mean anomaly a independent variable. In such a way, they obtain that the right-hand sides of the Lagrange equations depend only on $t$, which allows \citet{Isma11} to integrate them over one orbital period. Notice that they do not keep the wave's harmonic function constant in the integration. Finally, they compute the long-term changes of\footnote{It is unclear if the figures in Table III of \citet{Isma11} refer to the rates of change of the Keplerian orbital elements, i.e. if they are m s$^{-1}$ and deg s$^{-1}$, or to their  shifts over one orbital period, i.e. if they are m and deg. Also the units adopted in Table III of \citet{Isma11} are unspecified.} $a,e,I,\Om,\omega$ of Venus and Pluto for three, unspecified, different sources of gravitational waves.
\section{Summary and conclusions}\lb{conclu}
We analytically worked out the long-term variations of all the six {osculating} Keplerian orbital elements of a test particle orbiting a central body due to the perturbing tidal acceleration induced by the passage through the system of a monochromatic plane gravitational wave. We assumed that its frequency is much smaller than the orbital frequency of the test particle, so that the  time-dependent harmonic functions of the wave were kept constant in the integrations performed over one orbital period. We {considered}  a generic direction of the wavevector.

All the osculating Keplerian orbital elements, apart from the semi-major axis, undergo long-term variations. In particular, the eccentricity changes, so that the overall appearance of the orbit is altered as well: for example, the perihelion and aphelion distances vary accordingly. Such long-term variations are linear combinations of the non-vanishing elements of the tidal matrix, containing the slowly time-varying harmonic terms due to the wave, with coefficients which are complicated functions of the eccentricity, the inclination, the node and the pericenter of the test particle. We did not make any a priori simplifying assumptions concerning the inclination and the eccentricity of the test particle's orbit; thus, we obtained exact results as far as such a point is concerned. Moreover, their validity is not restricted to any specific reference frame. The variation of the eccentricity is proportional to the eccentricity itself, so that circular orbits do not change their shape. In the case of incidence normal to the orbital plane, also the inclination is left unaffected.

From a practical point of view, in the most general case one has six unknowns: the two angles of the wavevector, the frequency of the wave, its two independent amplitudes, and a phase lag. In principle, it is possible to constrain all of them by determining the secular variations of some Keplerian orbital elements for, e.g., several planets of the solar system. Actually, this seems to be the current trend in astronomical research since extensive planetary data reductions, in which the corrections to the standard precessions of the perihelia and the nodes of some major bodies of the solar system are estimated,  are currently ongoing by independent teams of astronomers. Future implementation of planned or proposed accurate interplanetary ranging facilities may further enhance such a program. It should also be considered that the resulting bounds would represent independent tests of the ultra-low frequency primordial gravitational waves which may have been indirectly detected in recent data analyses of BICEP2. A systematic analysis of the currently available planetary data for various possible orientations of the wavevector may be the subject of further, dedicated investigations.

Our results are quite general since they are valid not only for the solar system's planets, but also for any generic gravitationally bound two-body systems whose characteristic orbital frequencies are quite larger than the incident wave's frequency. {Typical examples are extrasolar planetary systems and the stars orbiting the Galactic black hole}. Moreover, {our findings} can be extended to any generic perturbing acceleration of tidal type with constant coefficients  over the characteristic timescales  of the perturbed system. {Exotic modified models of gravity and/or various types of external cosmological curved backgrounds may do so}.


\end{document}